\documentclass[letterpaper,prl,twocolumn]{revtex4}
\usepackage{graphicx,bm}
\usepackage{amsmath,amssymb}

\begin{document}

\title{Fluctuation-induced dissipation in evolutionary dynamics}
\author{Tsung-Cheng Lu$^{1}$, Yi-Ko Chen$^{1}$, Hsiu-Hau Lin$^1$ and Chun-Chung-Chen$^2$}
\affiliation{
$^1$Department of Physics, National Tsing Hua University,
30013 Hsinchu, Taiwan\\
$^2$ Institute of Physics, Academia Sinica, Nankang, Taipei 11529, Taiwan
\vspace{2mm}\\
Correspondence and requests for materials should be addressed to\\
H.-H. L. (hsiuhau.lin@gmail.com)
or C.-C. C (cjj@phys.sinica.edu.tw)
}
\date{October 4, 2014}

\begin{abstract}
Biodiversity and extinction are central issues in evolution. Dynamical balance among different species in ecosystems is often described by deterministic replicator equations with moderate success. However, fluctuations are inevitable, either caused by external environment or intrinsic random competitions in finite populations, and the evolutionary dynamics is stochastic in nature. Here we show that, after appropriate	coarse-graining, random fluctuations generate dissipation towards extinction because the evolution trajectories in the phase space of all competing species possess positive curvature. As a demonstrating example, we compare the fluctuation-induced dissipative dynamics in Lotka-Volterra model with numerical simulations and find impressive agreement. Our finding is closely related to the fluctuation-dissipation theorem in statistical mechanics but the marked difference is the non-equilibrium essence of the generic evolutionary dynamics. As the evolving ecosystems are far from equilibrium, the relation between fluctuations and dissipations is often complicated and dependent on microscopic details. It is thus remarkable that the generic positivity of the trajectory curvature warrants dissipation arisen from the seemingly harmless fluctuations. The unexpected dissipative dynamics is beyond the reach of conventional replicator equations and plays a crucial role in investigating the biodiversity in ecosystems.
\end{abstract}

\maketitle

Biodiversity is commonly used to indicate the stability of an ecosystem\cite{May74,Pimm84,Jablonski08}.
One of the central issues is to effectively promote the biodiversity while attracting more scientists' attention from various fields\cite{McLaughlin02,Both06,Sala00, Reichenbach07, Loreau01}.
The causes that threaten the biodiversity, for instance, climate change\cite{McLaughlin02,Both06}, over-harvesting, habitat destruction\cite{Sala00}, and population mobility\cite{Reichenbach07}, are well studied.
Above those factors, Darwin's theory of natural selection plays a crucial role in catalysis\cite{Smith82,Hofbauer98,Nowak06,Nowak06a}.
People are warned to reduce these effects in order to maintain and reserve the nature's biodiversity.
Nevertheless, a naive reversed statement should be check, namely, would the ecosystem be perfectly stable without any of these hazardous factors?

To put the discussions on firm ground, we can start with the non-transitive rock-paper-scissors game\cite{Drossel01,Kerr02,Czaran02,Nowak04,West06,Szabo07}, known as a paradigm to illustrate the species diversity.
When three subpopulations interact in this non-transitive way, we expect that each species can invade another when its population is rare but becomes vulnerable to the other species when over populated.
The non-hierarchical competition\cite{Traulsen05,Frey08,Traulsen08,Frey09} gives rise to the endlessly spinning wheel of species chasing species and the biodiversity of the ecosystem reaches a stable dynamical balance.
This cyclic evolutionary dynamics has been found in plenty of ecosystems such as coral reef invertebrates\cite{Jackson75}, lizards in the inner Coast Range of California\cite{Sinervo96} and three strains of colicinogenic \textit{Escherichia coli}\cite{Kerr02,Kirkup04} in Petri dish.
Although the oscillatory solutions for the replicator equations capture the main features, inclusion of mobility\cite{Reichenbach07} or/and finite-population effects\cite{Traulsen08,Frey09} in the numerical simulations always jeopardizes the stable equilibrium and highlight the importance of stochasticity in the evolution.

Populations in an ecosystem are discrete integers.
Approximating these discrete populations by continuous variables inevitably introduces \textit{intrinsic fluctuations}, which turn the evolutionary dynamics stochastic in nature.
By extensive numerical simulations, we record how the biodiversity of the ecosystem dissipates from these intrinsic fluctuations. In addition to the irregular deviations at the short-time scale, slowly but surely, dissipation induced from intrinsic fluctuations is derived and lead to an extinction time proportional to the population size\cite{Ifti03,Reichenbach06}.
Our findings can be elegantly summarized in three steps: discreteness induces fluctuations, fluctuations spawn dissipations and dissipative dynamics leads to extinction.

\section{Results}

\begin{figure}
\centering
\includegraphics[width=5cm]{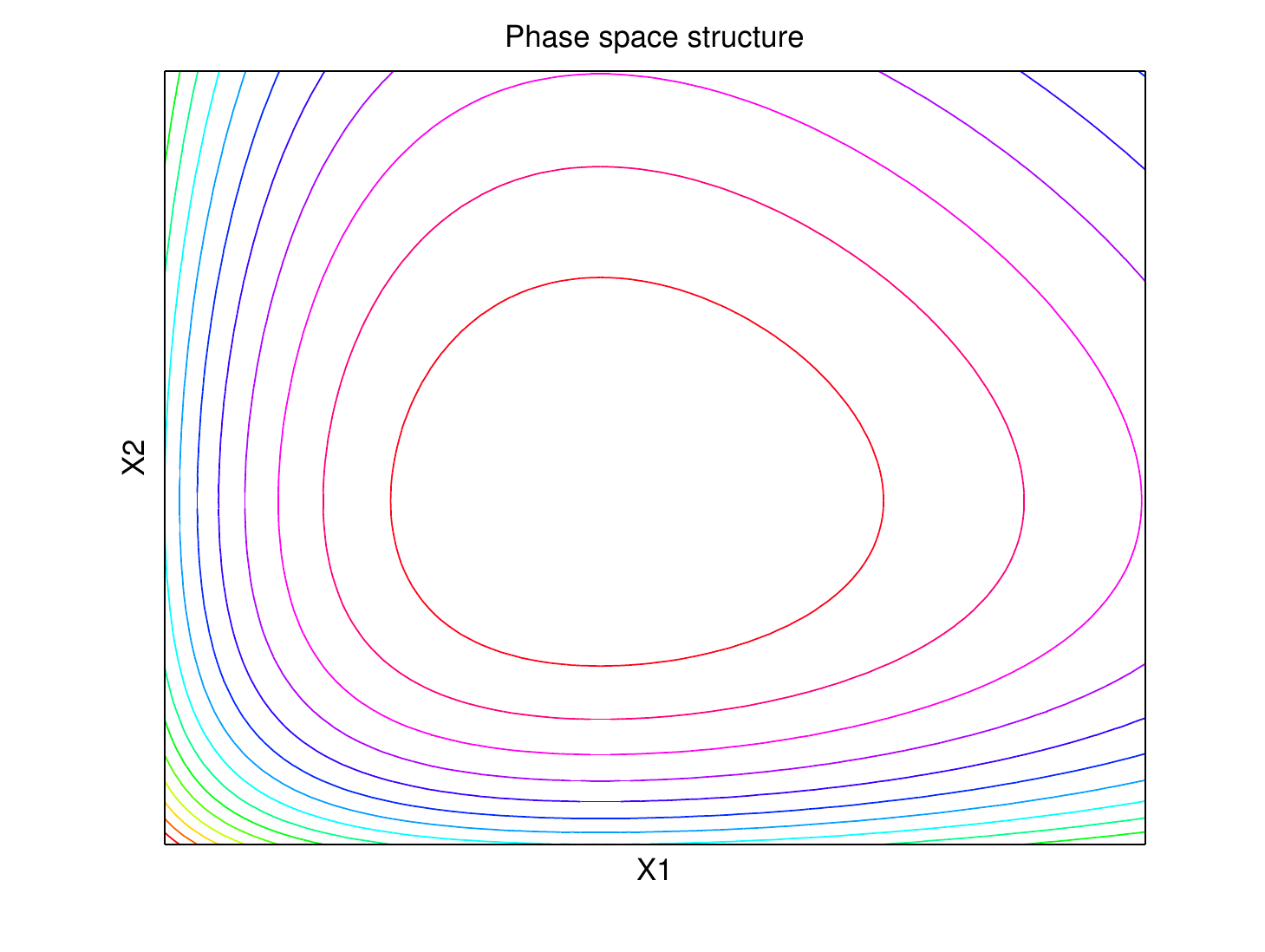}
\includegraphics[width=3cm]{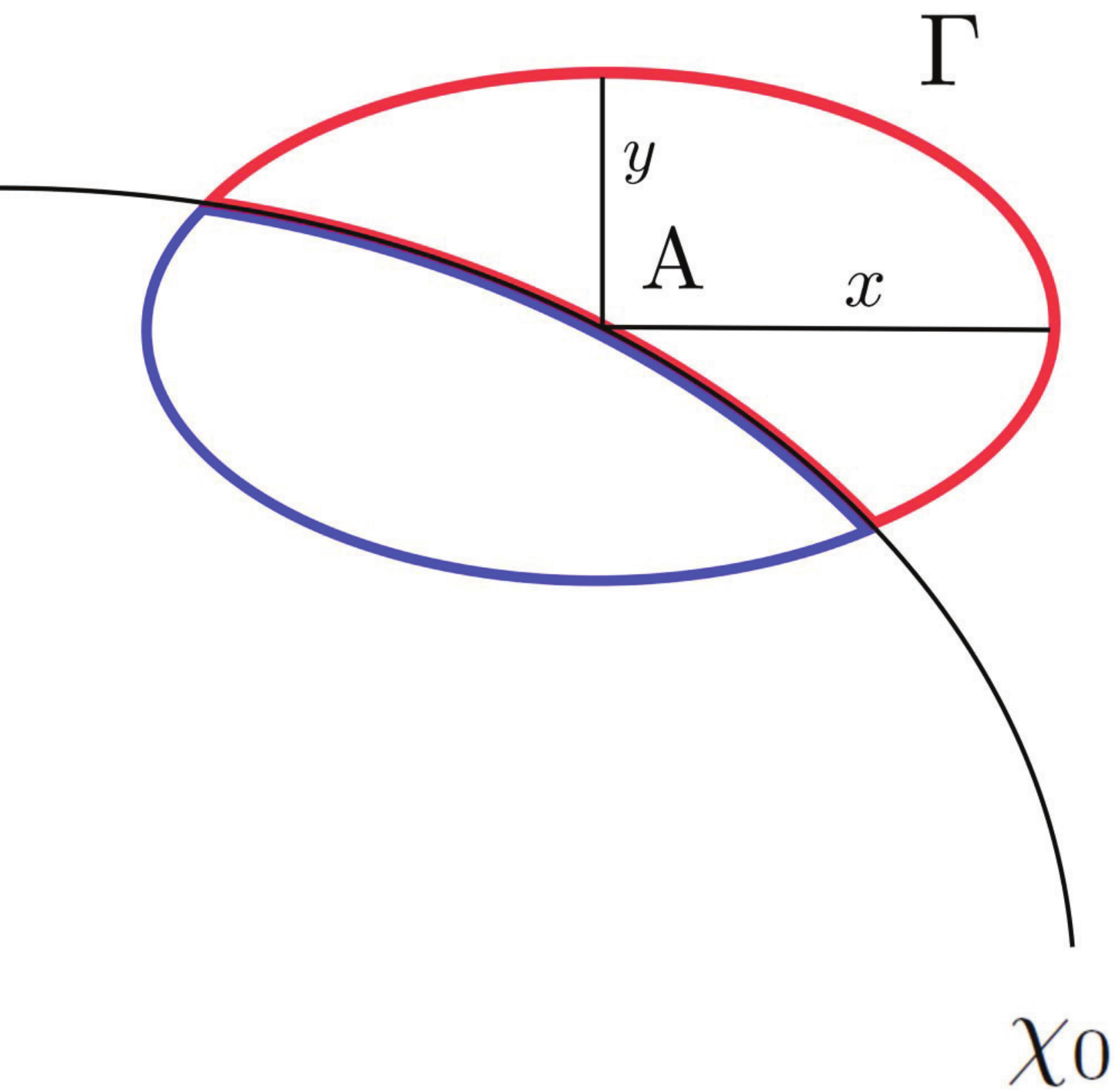}
\caption{Evolutionary contours in the phase space and the enlarged segment with local coordinates.}\label{contours}
\end{figure}

\noindent\textbf{Intrinsic fluctuations.} We start with the simplest case with two species A and B, whose population numbers are denoted by $x_1$ and $x_2$ respectively. Because the change of the population number of species is always discrete, the phase space spanned by $x_1$ and $x_2$ has internal uniform grid structure, where the system can only evolve on grid points. For a given state of system on a grid point in a small time step, the system can either hop to adjacent point or stay at the same point. Therefore, the system is stochastic in nature and has to be described by probability which is determined by competing relation among species. As long as the total population $N$ is large, with an appropriate coarse graining, we can derive effective replicator equations with details found in the Methods section,
\begin{equation}
\label{eq:stochastic eq}
\frac{dx_i}{dt}=f_ix_i+\xi_i  \qquad i=(1,2)
\end{equation}
where $f_i$ stands for the fitness which governs the deterministic dynamics in the system, and $\xi_i$ is the intrinsic noise arisen from the discreteness of population. Note that $\xi_i$ approximates to Gaussian noise under the large-N limit. If deterministic dynamics of system driven by $f_ix_i$ predicts a closed trajectory in the phase space, there exists a constant of motion which is determined by the initial condition  $(x_1(0),x_2(0))$, and thus the system can evolve on different contours depending on different initial condition(Fig.\ref{contours}). However, system will deviate from the original orbit due the random noises. To capture the influence caused by the noises, we can define stability indicator $\chi$, which is a conserved quantity within the deterministic dynamics, and let $\chi $ be the maximum at the fixed point predicted by the deterministic dynamics. With this indicator, we can measure the effect of fluctuations by studying the time evolution of $\chi$.

To realize the influence of noise, let us focus on the system at point A on the orbit as shown in Fig. 1. During a small time interval $\Delta t$, the system will have a stochastic change in the direction of $x_1$ and $x_2$ caused by $\xi_1$ and $\xi_2$ respectively. If we mark all the probable points with the same probability that the system can arrive in $\Delta t$, then we will get a closed orbit $\Gamma$ which shows reflectional symmetry with the vertical and horizontal axis with respect to point A due to the isotropy of noise under large-N limit. Due to the curvature of equi-$\chi $ contour, the area surrounded by blue line corresponding to $\chi > \chi _0$ is greater than the area surrounded by red line corresponding to $\chi <\chi_0$. Therefore, the expectation value of the change of $\chi$ will be negative(i.e.$\left<\Delta \chi\right><0 )$, which indicates that the dissipation of $\chi$ depends on the curvature of equi-$\chi $ contours. Because the ratio of these two areas will increase when the curvature increases, we can imagine that the magnitude of dissipation of $\chi $ and local curvature is positive correlated. Besides, for a given distance that system can travel, $\Delta \chi$ is larger in the area with denser equi-$\chi $ contours, which is a reflection of the magnitude of gradient $|\nabla \chi|$. Under large-N limit, for the case that the standard deviations of Gaussian noises $\sigma$ in both vertical and horizontal directions are the same, the probable hopping points correspond to equal probability becomes a circle centered at point A. If we take the small curvature approximation $\kappa\ll\frac{1}{\sigma}$ and assume that local curvature of contours $\kappa$ and $|\nabla \chi|$ are constant, the formula for $\left<\Delta \chi\right>$ can be derived explicitly:
\begin{equation}
\label{eq: first moment}
\left<\Delta \chi\right>=-\frac{1}{2}\kappa |\nabla \chi | \sigma ^2
\end{equation}
The most important feature in Eq.(\ref{eq: first moment}) is that the magnitude of dissipation of $\chi $ has both linear dependance on curvature and gradient, which reveals that the dissipation is determined by the geometric structure of contours in the phase space. Additionally, the magnitude is proportional to $\sigma ^2$, which reflects the fact the dissipation occurs due to the fluctuations in the system. From the argument above, for those system with convex contours in phase space, $\left<\Delta \chi\right> $ is always negative, and thus the system is inclined to move to outer and outer contours. When the system arrive at the boundaries, only one surviving species is left, revealing the destruction of biodiversity. Practically speaking, because any kind of species cannot survive without the interaction with other species, the ecosystem is destined to go to extinction.

\begin{figure}
\centering
\includegraphics[width=8cm]{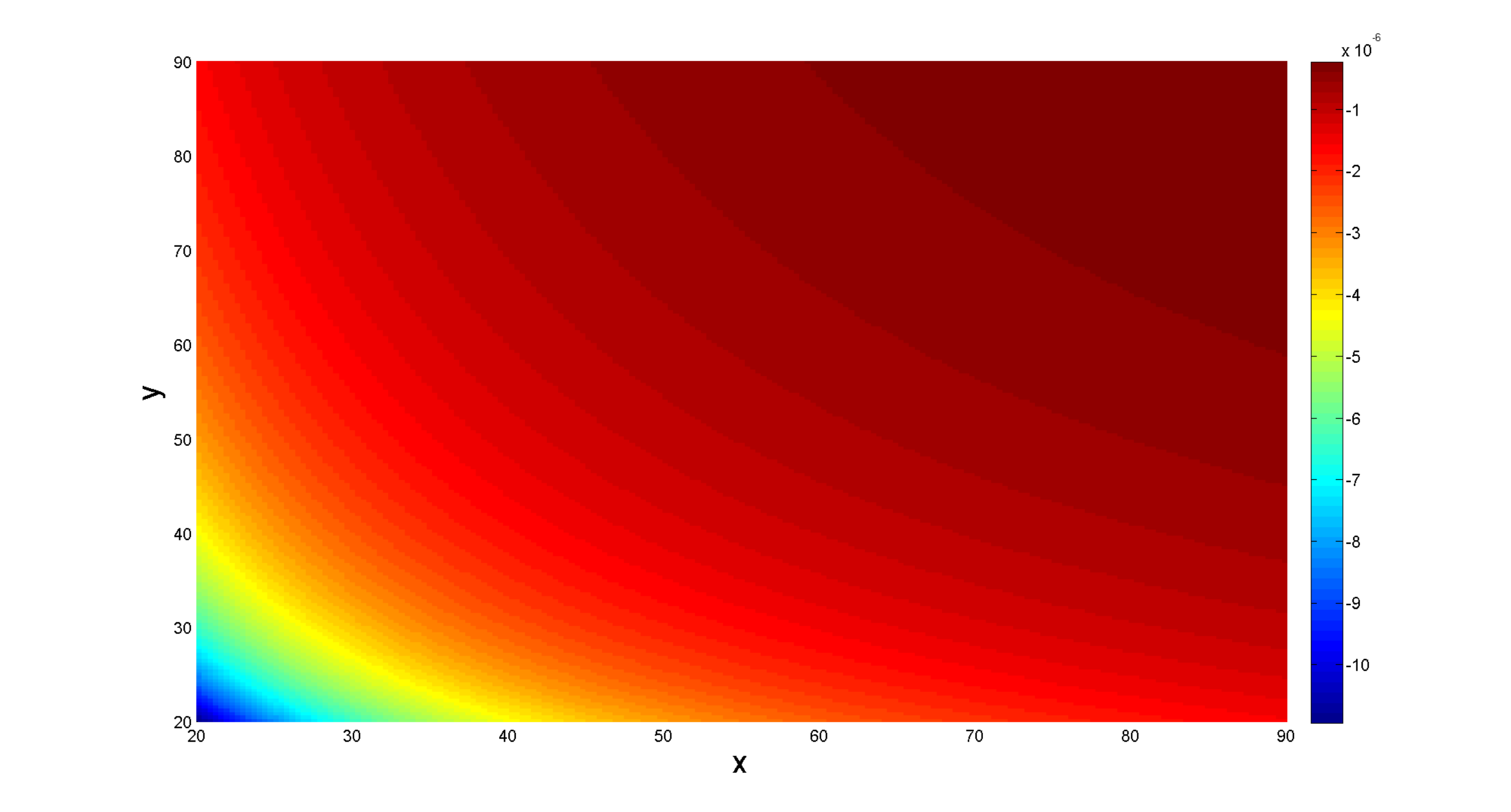}
\includegraphics[width=8cm]{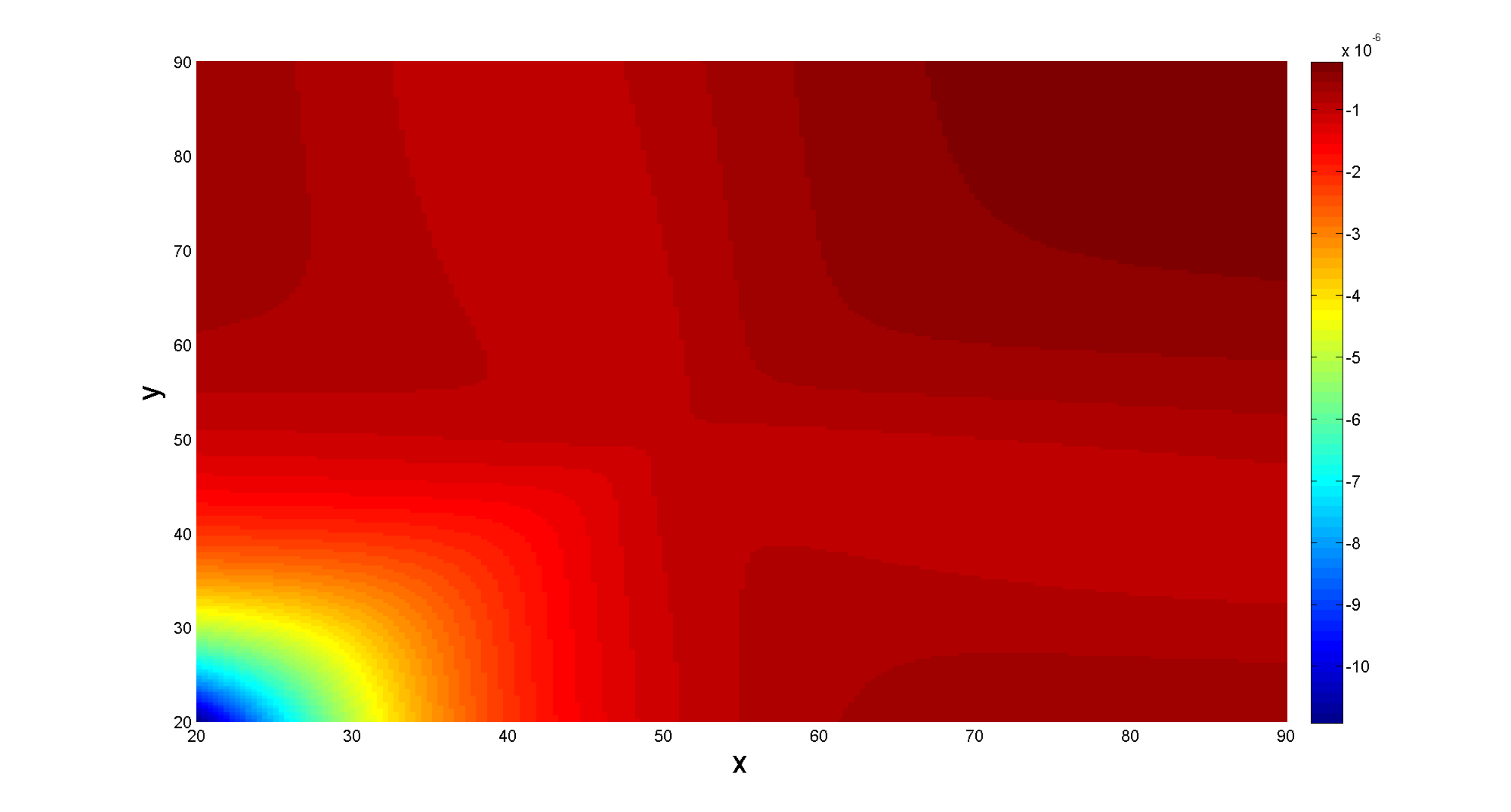}
\caption{
(a)$\Delta \chi$ calculated by algorism in phase space with each species number ranges from 20 to 90. (b)$\Delta \chi $ calculated by the theory in phase space with each species number ranges from 20 to 90.
}
\end{figure}

The above method can be easily generalized to N-species stochastic system, where N can be any finite positive integer. In N-dimensional phase space, if the deterministic dynamics predicts a closed convex orbit, where we can define a biodiversity indicator $\chi $ for a orbit, the structure of phase space can be characterized by a hypersurface for different value of $\chi $. During a small time interval $\Delta t$, for a given state on a contour, we can plot a hypersurface $\Gamma$ which is the set of equally probable points that the system can hop into. Due to the convexity of equi-$\chi $ hypersurface, the volume closed by the hypersurface $\Gamma$ is divided to two parts with different volumes. In consequence, the system is inclined to go to outer hypersurface, implying the destined extinction of the system. Therefore, for any dimensional phase space where deterministic driving force predicts a closed topological structure, the seemly harmless fluctuation can always induce the dissipation, and thus drive the system to extinction.

\noindent\textbf{Lotka-Volterra Model.}
We present Lotka-Volterra model as a demonstration of our method. The Lotka-Volterra model expresses the dynamics of the prey and the predator in an ecosystem. Eqs.(7) are the replicator equation, with with x and y denoting species population of X(predator) and Y(prey) respectively. Besides the interaction part, there are natural birth of X and natural death of Y using coefficient a and c in the mechanism.

\begin{align}
\frac{dx}{dt} &= -ax + bxy
\nonumber\\
\frac{dy}{dt} &= cy - dxy
\end{align}

It is clear that the populations of prey and predator both present stably oscillatory motions with respect to time, which is equivalent to closed orbits in the phase space. Using this deterministic approach, there is a stability indicator $\chi$ corresponding to each orbit(Fig.(3)).
\begin{equation}\label{2}
  \chi =-by-dx+a \log y+c \log x
\end{equation}
However, the replicator equations couldn't capture what truly happens for a finite population system. There are always
intrinsic fluctuations, which can not be neglected, due to finite population. What is the outcome of this stochastic behavior for Lotka-Volterra model? In Fig.(3), we observe closed circles with positive curvature in phase space. Therefore, from our method, when noise is presented in this situation, dissipative motion would appear. The system will deviate away from its original contour and finally go to extinction because of absorbing boundaries in phase space. In order to see how the method agrees with realistic situation, we calculate the first moment of $\chi$, which indicates dissipation, in phase space. The algorism for the stochastic motion is to use probability to describe the mechanism. Not surprisingly, $\chi$ is negative over the whole phase space. Fig.(4)shows great agreement between the algorism and the theoretical calculation in region A of Fig.(3). In region A, it's natural to have successful agreement because that all the approximations we have done during calculation are satisfied. During the derivation of the method, there are two assumptions we have used. One is that the standard deviation of Gaussian distribution of noise is much smaller than the radius of curvature of contour. The other is that during the hopping of species from one point to another in the phase space, gradient of stability indicator and curvature of contour remain the same within the range of the standard deviation. In region A, the curvature of contour is small compared to other areas and the change of curvature and of gradient is also small. Therefor, a good demonstration between algorism and theoretical calculation is presented in this area. However, the conditions in region B in Fig.(3) are far away from the approximations and hence conflictions appear.

The neutral coexistence predicted by deterministic approach is broken due to the intrinsic fluctuation of a biological system. The fluctuation-induced dissipation is generated and we use the first moment of $\chi$ to capture the dissipative behavior. Successfully, the great agreement between the algorism and the theoretical calculation in the area we focus on is found in Lotka-Volterra model and this presents the validity of the method.

\noindent\textbf{summary.} We present a method that can concretely predict the fluctuation-induced dissipation in a biological system. Although the deterministic approach illustrates the beautiful coexistence, this method ignores the unavoidable noise in the evolution due to the external influence of environment or the intrinsic fluctuation of a finite population system. As a result, it is better to use stochastic method to describe evolution; however, the extinction of the species always occur in this kind of simulation. To have a more profound understanding of the extinction process, we take an appropriate coarse-graining of phase space structure and define a stability indicator $\chi$ to describe the extinction process of system, finding that the fluctuations always induce the dissipation by the positive curvature of the equi-$\chi $ contour either in neutral stable or stable cases. As a demonstrating example, we compare theoretical calculation and simulation of stochastic Lotka-Volterra model and find fantastic agreement. Furthermore, our finding shares great similarity with fluctuation-dissipation theorem in statistical physics  while the major difference is that our method is valid even in the system which is far from equilibrium. The breakdown of biological diversity in ecosystem is not so mysterious. From the geometric structure of the stability indicator $\chi $ in phase space, we can see that the fluctuation-induced dissipation is predicted and plays an important role in the evolutionary dynamics for biological systems.

\section{Discussion}
The spirit of the our idea is that fluctuation can induce a non-zero dissipation through the geometric structure of the phase space determined by the deterministic dynamics. Therefore, the idea is close related to the fluctuation-dissipation theorem in statistical physics. To further illustrate the similarity, for two-species system, under the the same approximation as in the calculation of the first moment, we can derive the formula for the second moment of stochastic variable $\chi$:
\begin{equation}
\label{eq: second moment}
\left<\Delta \chi ^2\right>=|\nabla \chi |^2 \sigma ^2
\end{equation}
which is a reasonable result because second moment reflects magnitude of fluctuation of the stochastic variable $\chi$ which depends on the density of contours and intensity of fluctuation only, instead of curvature of phase contours.

Take the continuous limit of time interval, the correlation of noise on $x_1$ and $x_2$ is $\left< \xi_i(t_1)\xi_i(t_2)  \right> =D\delta (t_1-t_2)$ ,$i=1,2$, then we can derive the generalized Langevin equation for the stochastic variable $\chi$:
\begin{equation}
\label{eq:langevin eq}
\frac{d\chi}{dt}=-\frac{1}{2}\kappa |\nabla \chi |D+\xi_\chi (t)
\end{equation}
where $\left< \xi_\chi(t_1)\xi_\chi(t_2) \right>=|\nabla \chi |^2 D\delta (t_1-t_2)$.
From this expression, it is obvious to see that the dissipative dynamics emerges from the seemingly harmless intrinsic fluctuations, and the connection of fluctuation and dissipation is in the same way as Fluctuation-dissipation theorem. However, the marked difference is that the relation we derive can apply to the system which is far from equilibrium while Fluctuation-dissipation theorem only works in the linear response regime of a system. Since the idea is valid for the system which is far from equilibrium, we may interpret our result as a generalization of Fluctuation-dissipation theorem for the ecosystem.

In the above discussion, we focus on a system where all species display dynamical balance with a neutral stability, and discover that the fluctuations induce a dissipation which drives species to extinction. However, most of dynamical systems show either stable or unstable stability instead of a neutral dynamical balance. While extinction in the unstable system is expected, the extinction process of a stable system is more complicated because it is hard to capture the effect caused by fluctuations. In order to investigate the effect of fluctuation on the stable system, consider a N-species system where there is only one stable fixed point. Although the deterministic flow in the phase space spanned by population numbers of species $x_i(i=1,2...N)$ drives the system toward the fixed point, we can naively think that extinction always occurs due to a series of unfortunate events as long as the evolution time is long enough. To measure the effect caused by fluctuations quantitatively, we can define stability indicator $\chi $ as the following:

\begin{equation}
\label{eq:stabilityindicator}
\chi(x_1,x_2,...x_N)=-\alpha\sum_{i=1}^N(x_i-x_i^*)^2
\end{equation}
where $(x_1^*,x_2^*,...x_N^*)$ denotes the coordinate of fixed point, and $\alpha$ is a positive constant. By this definition, the set of equi-$\chi$ points forms a closed hypersurface which centers at the fixed point, where $\chi$ achieves maximum. Therefore, we can measure the extinction process by studying the dissipation of $\chi$, which indicates the evolution of the system toward the boundaries. Ignore the noise first, different from the case of neutral dynamical system, the deterministic flows contribute to the time evolution of $\chi$ as well. For the effect of fluctuations, because the equi-$\chi$ hypersurface is closed, we can still apply our previous method to determine the dissipation caused by fluctuations. Again, during a small time interval, the set of points with the same probabilities that system can hop on forms a closed hypersurface. With the positive curvature of equi-$\chi$ hypersurface, there is an asymmetry of phase space volume between inner and outer region, and thus the fluctuation-induced dissipation is expected.
Furthermore, we can derive the Langevin equation of $\chi$:
\begin{equation}
\label{eq:langevin eq}
\frac{d\chi}{dt}=\nabla \chi\cdot (\frac{d\vec{x}}{dt})_{flow} -\frac{1}{2}\kappa |\nabla \chi |D+\xi_\chi (t)
\end{equation}
where $\left< \xi_\chi(t_1)\xi_\chi(t_2) \right>=|\nabla \chi |^2 D\delta (t_1-t_2)$, and $(\frac{d\vec{x}}{dt})_{flow}$ denotes the deterministic flow in the phase space which always increases the value of $\chi$. Therefore, the extinction process of species is complicated due to the competition between deterministic flow and fluctuation-induced dissipation. However, due to the existence of white noise, the probability distribution in the $\chi$ will spread throughout the $\chi $ space. In consequence, the system will still go to extinction due to the absorbing boundaries of phase space spanned by N-species for a large enough time.

\section{Methods}

\noindent\textbf{Differential equations in continuous limit.}
The most fundamental description for biological evolution is to depict the discrete change of species population with probabilities. From this microscopic view of how the species evolves, Eq.(1) can be derived. To start with, we take the Lotka-Volterra model, with two species $X_{1}$ and $X_{2}$, as an example. For species $X_{1}$, during the simulation time interval $\Delta \tau$, where $\tau = 1,2,3,...$, the population change $\Delta x_{1}$ may be $1$, $0$ or $-1$, which corresponds to the probability $P_{+1}$, $P_{0}$ and $P_{-1}$ respectively. We can write down the average change of $x_{1}$ during a simulation time interval.
\begin{eqnarray}
\langle \Delta x_{1} \rangle &=& 1\cdot P_{+1}+0\cdot P_{0}+(-1)\cdot P_{-1}
\nonumber\\
&=& ax_{1}\Delta t-bx_{2}x_{1}\Delta t
\end{eqnarray}
Probability can be naturally defined as $P_{\alpha}=A_{\alpha}x_{1}\Delta t$, where $A_{\alpha}$ depends on evolution condition. For example, $P_{+1}=a$ and $P_{-1}=bx_{2}$. Besides, to make the value of probability smaller than one, we set the natural time interval $\Delta t=\Delta \tau/N^{2}=1/N^{2}$ in this case. Since Eq.(18) is the average value, to have the explicit expression of $\Delta x_{1}$ during $\Delta \tau$, the noise must be added.
\begin{equation}
\Delta x_{1}  = \langle \Delta x_{1} \rangle + \eta_{1}
\end{equation}
Next, we devide by $\Delta \tau$ and use the property $\Delta \tau=1$:
\begin{equation}
\frac{\Delta x_{1}}{\Delta \tau}  = \frac{\langle \Delta x_{1} \rangle}{\Delta \tau} + \frac{\eta_{1}}{\Delta \tau}
= \langle \Delta x_{1} \rangle + \eta_{1}
\end{equation}
Using the relation $\Delta \tau=N^{2}\Delta t$, we replace $\Delta \tau$ in the left hand side with $N^{2}\Delta t$ and move $N^{2}$ to the right. Therefore, we have the following equation:
\begin{eqnarray}
\frac{d x_{1}}{d t}  \simeq
\frac{\Delta x_{1}}{\Delta t} &=&  \langle \Delta x_{1} \rangle N^{2} + \eta_{1} N^{2}
\nonumber\\
&=& ax_{1}-bx_{2}x_{1}+\xi_{1}
\end{eqnarray}
When populatin number $N$ is large enough, $\Delta t$ would be limited to zero and hence we can approximate the discrete population change to continuous differential equation. If Eq.(21)is taken the average, the noise term would disappear and then turn back to the replicator equation. The similar derivation could be applied to multiple species. As a result, we have the generalized version Eq.(1)in the first place. This part of demonstration shows that the stochastic differential Eq.(1) is derived from the microscopic view of how species interacts with others. With large population $N$ approximation, we have the differential equation which comes from the discrete stochastic equation.

\begin{figure}
  \centering
  \includegraphics[scale=0.3]{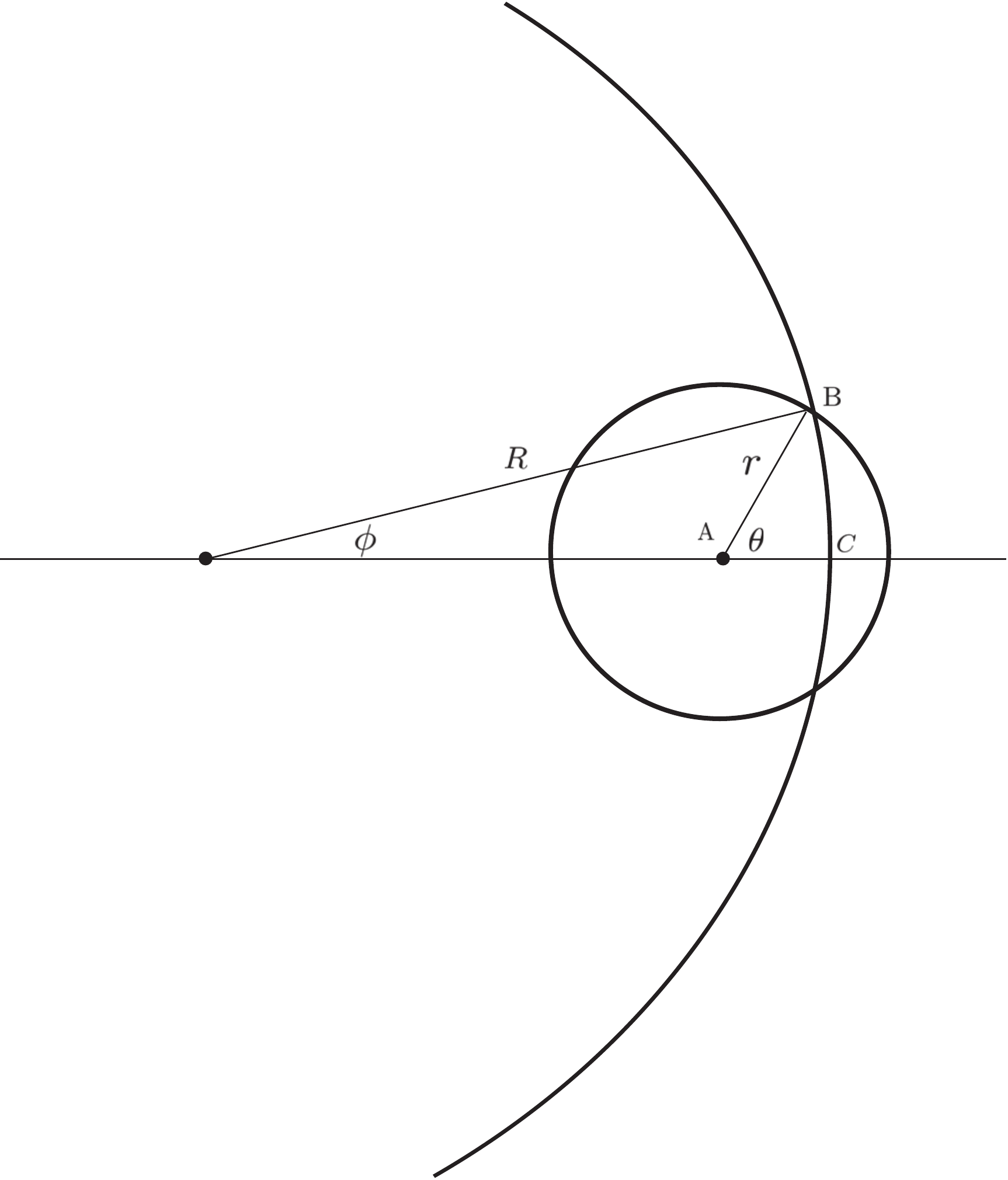}\\
  \caption{Geometric definitions for the angular variables used in the derivations.}\label{supplementary2}
\end{figure}

\noindent\textbf{Dissipation in biodiversity indicator.}
Here we present the details for the calculation of $\left<\Delta \chi\right>$ and $\left<\Delta \chi ^2\right>$: \\
     For system located at point A as shown in (Fig.\ref{supplementary2}), if we mark all the probable points that system can hop within small time interval $\Delta t$, the set of points will form a circle centered at point A under the approximation that random white noises obey Gaussian distribution with equal standard deviation. Inside the circle with radius $r$, assuming $|\nabla \chi |$ and $\kappa$ are constant, we can calculate the change of $\chi$ corresponding to different hopping points$\Delta \chi(r,\theta)$:
\begin{equation}
\label{eq: delta chi}
\begin{aligned}
\Delta \chi (r,\theta) =-|\nabla \chi |\overline{AC}=-|\nabla \chi |(R-R\cos\phi +r\cos\theta)
\end{aligned}
\end{equation}
Use the distance between point B and $\overline{AC}$ to find the relation between $\phi$ and $\theta$
\begin{equation}
\label{eq: relation between two angle}
\begin{aligned}
R\sin\phi =r\sin\theta\\
\cos\phi=\frac{\sqrt{R^2-r^2\sin^2\theta}}{R}
\end{aligned}
\end{equation}
Substituting Eq.(\ref{eq: relation between two angle})to Eq.(\ref{eq: delta chi}):
\begin{equation}
\begin{aligned}
\Delta \chi =-|\nabla \chi |(R-\sqrt{R^2-r^2\sin^2\theta}+r\cos\theta)
\end{aligned}
\end{equation}
Taking the angular average of $\Delta \chi$:
\begin{equation}
\label{eq: angular average}
\begin{aligned}
\left<\Delta \chi\right>_\theta &=\frac{1}{2\pi}\int_0^{2\pi} \Delta \chi d \theta\\
& \hspace{-1cm} =\frac{1}{2\pi}\int_0^{2\pi}-|\nabla \chi |(R-\sqrt{R^2-r^2\sin^2\theta}
+r \cos\theta)d\theta \\
& \hspace{-1cm} =-\frac{|\nabla \chi|}{2\pi}\left[2\pi R-R\int_0^{2\pi} 
\sqrt{1-(\frac{r\sin\theta}{R})^2 }d\theta \right]
\end{aligned}
\end{equation}
under the approximation that $r\ll R$, we can evaluate the integral by binomial expansion to first order:
\begin{equation}
\label{eq: binomial expansion}
\begin{aligned}
\int_0^{2\pi} \sqrt{1-(\frac{r\sin\theta}{R})^2 }d\theta
&=\int_0^{2\pi} \left(1-\frac{r^2}{2R^2}\sin^2\theta\right)d\theta\\
&=2\pi-\frac{|\nabla \chi| r^2}{4R}
\end{aligned}
\end{equation}
Substituting Eq.(\ref{eq: binomial expansion}) to Eq.(\ref{eq: angular average}):

\begin{equation}
\label{eq:first moment theta}
\begin{aligned}
\left<\Delta \chi\right>_\theta =-\frac{|\nabla \chi| r^2}{4R}
=-\frac{1}{4}\kappa |\nabla \chi| r^2
\end{aligned}
\end{equation}
Following similar steps, procedure, we can calculate
\begin{equation}
\label{eq:second moment theta}
\begin{aligned}
\left<\Delta \chi^2\right>_\theta =\left<\Delta \chi\right>_\theta =\frac{1}{2\pi}\int_0^{2\pi} d\theta \Delta \chi^2
=\frac{1}{2}|\nabla \chi|^2 r^2
\end{aligned}
\end{equation}
From Eq.(\ref{eq:first moment theta}) and Eq.(\ref{eq:second moment theta}), we can already observe that the first moment has linear dependence on curvature and gradient $\chi$ while the second moment depends on gradient only. To obtain explicit result,
assuming the random displacement$(x,y)$ caused by random white noises obey Gaussian distribution,
\begin{equation}
\begin{aligned}
P_x(x)=\frac{1}{\sqrt{2\pi\sigma ^2}}e^{-\frac{x^2}{2\sigma^2}}\\
P_y(y)=\frac{1}{\sqrt{2\pi\sigma ^2}}e^{-\frac{y^2}{2\sigma^2}}.
\end{aligned}
\end{equation}
It is straightforward to compute the radial distribution of noises,
\begin{equation}
\begin{aligned}
P_r(r)=\frac{r}{\sigma^2}e^{-\frac{r^2}{2\sigma^2}}
\end{aligned}
\end{equation}
where $S$ refers to the area in the circle with radius r.
Assuming $\sigma\ll R$, due to extreme narrow peak of probability distribution, we can safely use Eq.(\ref{eq:first moment theta}) in the calculation of radial average of $\left<\Delta \chi\right>_\theta$ even for $r\rightarrow \infty$ in the integral . Thus we can evaluate $\Delta \chi$ by taking radial average of $\left<\Delta \chi\right>_\theta$ from the contributions of different hopping radius:
\begin{equation}
\label{eq:explicite result1}
\begin{aligned}
\left<\Delta \chi\right> &=\int_0^\infty dr Pr(r)\left<\Delta \chi\right>_\theta
\\
&=-\frac{\kappa}{4\sigma^2}|\nabla \chi|\int_0^\infty r^3e^{-\frac{r^2}{2\sigma^2}}dr
\\
&=-\frac{1}{2}\kappa \sigma^2|\nabla \chi|
\end{aligned}
\end{equation}
Similarly, the average of the second moment can also be obtained,
\begin{equation}
\label{eq:explicit result 2}
\left<\Delta \chi ^2\right>=|\nabla \chi |^2 \sigma^2.
\end{equation}

\section{Acknowledgements}
We acknowledge supports from the Ministry of Science and Technology in Taiwan through grant MOST 103-2112-M-007-011-MY3. Financial supports and friendly environment provided by the National Center for Theoretical Sciences in Taiwan are also greatly appreciated.

\section{Author contributions}
T.C.L. and Y.K.C. performed the analytical and numerical calculations. H.H.L. and C.C.C. supervise the whole work. All authors contributed to the preparation of this manuscript.


\begin{thebibliography}{99}
\bibitem{May74}
R. M. May,
\textit{Stability and Complexity in Model Ecosystems}
(Princeton University Press, Princeton, New Jersey, 1974), second edition.

\bibitem{Pimm84}
S. L. Pimm,
Nature \textbf{307}, 321 (1984).

\bibitem{Jablonski08}
D. Jablonski
Proc. Natl. Acad. Sci. USA {\bf 105}, 11528 (2008).

\bibitem{McLaughlin02}
J. F. McLaughlin, J. J. Hellmann, C. L. Boggs, and P. R. Ehrlich,
Proc. Natl Acad. Sci. USA \textbf{99}, 6070 (2002).

\bibitem{Both06}
C. Both, S. Bouwhuis, C. M. Lessells, and M. E. Visser,
Nature \textbf{441}, 81 (2006).

\bibitem{Sala00}
O. E. Sala \textit{et al.},
Science \textbf{287}, 1770 (2000)

\bibitem{Reichenbach07}
T. Reichenbach, M. Mobilia and E. Frey,
Nature \textbf{448}, 1046 (2007).

\bibitem{Loreau01}
M. Loreau \textit{et al.},
Science \textbf{294}, 804 (2001).

\bibitem{Smith82}
J. M. Smith,
\textit{Evolution and the Theory of Games}
(Cambridge Univ. Press, Cambridge, 1982).

\bibitem{Hofbauer98}
J. Hofbauer and K. Sigmund,
\textit{Evolutionary Games and Population Dynamics}
(Cambridge Univ. Press, Cambridge, 1998).

\bibitem{Nowak06}
M. A. Nowak,
\textit{Evolutionary Dynamics}
(Belknap Press, Cambridge, Massachusetts, 2006).

\bibitem{Nowak06a}
M. A. Nowak,
Science {\bf 314}, 1560 (2006).

\bibitem{Drossel01}
B. Drossel,
Adv. Phys. \textbf{50}, 209 (2001).

\bibitem{Kerr02}
B. Kerr, M. A. Riley, M. W. Feldma and B. J. M. Bohannan, 
Nature {\bf 418}, 171 (2002).

\bibitem{Czaran02}
T. L. Czaran, R. F. Hoekstra and L. Pagie,
Proc. Natl Acad. Sci. USA \textbf{99}, 786 (2002).

\bibitem{Nowak04}
M. A. Nowak and K. Sigmund,
Science {\bf 303}, 793 (2004).

\bibitem{West06}
S. A. West, A. S. Griffin, A. Gardner and S. P. Diggle,
Nature Rev. Micro. {\bf 4}, 597 (2006).

\bibitem{Szabo07}
G. Szabo, and G. Fath,
Phys. Rep. \textbf{446}, 97 (2007).

\bibitem{Traulsen05}
A. Traulsen, J. C. Claussen and C. Hauert,
Phys. Rev. Lett. {\bf 95}, 238701 (2005).

\bibitem{Frey08}
T. Reichenbach and E. Frey,
Phys. Rev. Lett. {\bf 101}, 058102 (2008).

\bibitem{Traulsen08}
J. C. Claussen and A. Traulsen,
Phys. Rev. Lett. {\bf 100}, 058104 (2008).

\bibitem{Frey09}
M. Berr, T. Reichenbach, M. Schottenloher and E. Frey,
Phys. Rev. Lett. {\bf 102}, 048102 (2009).

\bibitem{Jackson75}
J. B. C. Jackson and L. Buss,
Proc. Natl Acad. Sci. USA \textbf{72}, 5160 (1975).

\bibitem{Sinervo96}
B. Sinervo and C. M. Lively,
Nature \textbf{380}, 240 (1996).

\bibitem{Kirkup04}
B. C. Kirkup and M. A. Riley,
Nature \textbf{428}, 412 (2004).

\bibitem{Ifti03}
M. Ifti and B. Bergersen,
Eur. Phys. J. E \textbf{10}, 241 (2003); Eur. Phys. J. B \textbf{37}, 101 (2004).

\bibitem{Reichenbach06}
T. Reichenbach, M. Mobilia, and E. Frey,
Phys. Rev. E \textbf{74}, 051907 (2006).

\end{thebibliography}
\end{document}